\documentclass[aps,prl,twocolumn]{revtex4}
\usepackage{graphicx,epsf}

\begin{document}

\title {
Spontaneous interlayer coherence in bilayer Kondo systems
}
\author{T. Senthil}
\affiliation{Department of Physics, Massachusetts Institute of
Technology, Cambridge MA 02139}
\author{Matthias Vojta}
\affiliation{\mbox{Institut f\"ur Theorie der Kondensierten Materie,
Universit\"at Karlsruhe, 76128 Karlsruhe, Germany}}

\date{March 8, 2005}

\begin{abstract}
Bilayer Kondo systems present interesting models to illustrate the competition between the
Kondo effect and intermoment exchange.
Such bilayers can exhibit two sharply distinct Fermi liquid phases which are distinguished
by whether or not the local moments participate in the Fermi sea.
We study these phases and the evolution from one to the other upon changing Kondo coupling.
We argue that an ordered state with spontaneous interlayer phase coherence
generically intervenes between the two Fermi liquids.
Such a condensate phase breaks a U(1) symmetry and is bounded by a finite-temperature
Kosterlitz-Thouless transition.
Based on general arguments and mean-field calculations
we investigate the phase diagram and associated quantum phase transitions.
\end{abstract}
\pacs{}

\maketitle


Landau's Fermi liquid theory provides the basis for our understanding of most conventional metals.
A particularly interesting application of Fermi liquid theory is to
intermetallic compounds containing localized spin moments on $d$ or $f$
orbitals and additional bands of conduction electrons.
These systems are conveniently modelled by the Kondo lattice
Hamiltonian, with exchange interactions between the
local moments and the conduction electrons, and possibly
additional exchange couplings between the local moments
themselves \cite{hewson}.
These display metallic ``heavy'' Fermi liquid phases
(albeit with large quasiparticle effective masses and other renormalizations)
in addition to  phases with magnetic or superconducting long-range order \cite{doniach}.
The quantum phase transitions between the various phases are currently subject of
intense research activities.
For instance the transition between the heavy Fermi liquid and a magnetic metal exhibits a
number of interesting non-Fermi liquid phenomena
which have not yet found a consistent theoretical explanation \cite{piers}.

Of particular interest in heavy electron physics is the issue of the topology of the
Fermi surface \cite{oshi}, and how it evolves across various phase boundaries.
It is believed by some workers in the field that this issue is at the heart of
the non-Fermi liquid physics observed near heavy electron critical points \cite{piers,edmft,ffl}.
Our understanding of such matters is rather primitive --
clearly it would help to build more intuition about such phenomena.
In this paper we consider a paradigmatic model that illustrates some of these issues
and the associated difficulties.

We consider bilayer Kondo systems, consisting of pairs of identical two-dimensional layers with local
moments and conduction electrons.
We assume that the conduction electrons cannot hop between the two layers.
As argued below, such bilayer systems allow
for two different paramagnetic Fermi liquid phases.
These two phases are sharply distinct: the distinction is in the topology
of the Fermi surface.
In one case the Fermi surface may loosely be dubbed ``large'' -- physically this happens
when the local moments participate in the Fermi surface.
In the other case the local moments are not part of the Fermi sea and the Fermi surface
may loosely be dubbed ``small''.
We emphasize that Luttinger's theorem \cite{oshi} is satisfied
in both phases -- however the Fermi surface topology is different,
implying that the two phases cannot be smoothly connected to each other.
We argue below that a novel ordered phase generically intervenes in between the two phases.
This ordered state involves spontaneous development of interlayer coherence despite
the absence of any direct single particle hopping.
The interlayer coherence may either be in a particle-hole (i.e exciton) condensate or
in a particle-particle condensate. The former is analogous to states much discussed in bilayer
quantum Hall structures \cite{qhth,qhexp} while the latter is a bilayer superconductor with Cooper pairs shared
between the two layers.
At finite temperatures above the ordered phase we expect signatures of critical
behavior associated to strong Fermi surface fluctuations.

To be specific, consider the Hamiltonian:
\begin{eqnarray}
H &=& -\!\!\sum_{\langle jj'\rangle\sigma\alpha} \!\!
t_{jj'} c_{j\sigma\alpha}^{\dagger} c_{j' \sigma\alpha} +
\frac{J_K}{2} \sum_{j\sigma\sigma'\alpha} \! \vec{S}_{j\alpha} \!\cdot c^\dagger_{j \sigma\alpha}
\vec{\tau}_{\sigma\sigma'} c_{j\sigma'\alpha} \nonumber \\
&&+ \sum_{\langle jj'\rangle\alpha} I_{jj'} \vec{S}_{j\alpha} \cdot \vec{S}_{j'\alpha}
+ \sum_{j} J_\perp \vec{S}_{j1} \cdot \vec{S}_{j2} .
\label{h}
\end{eqnarray}
The lattice sites are labelled with $j$ in each layer, and $\alpha=1,2$ is
the layer index. The local moments $\vec{S}_{j\alpha}$ are $S=1/2$ spins, and the
conduction electrons $c_{j \sigma\alpha}$ ($\sigma = \uparrow
\downarrow$) hop on the sites $j$, $j'$ of some regular lattice in
$d$ spatial dimensions ($d=2$ for most what follows)
with amplitude $t_{jj'}$.
$J_K >0$ are the Kondo exchanges, $I_{jj'}$ denote explicit short-range
Heisenberg interactions within each plane of local moments, and
$J_\perp$ is the interlayer exchange interaction.
A chemical potential for the $c_\sigma$
fermions which fixes their density at $\rho_c$ per unit cell of
the ground state is implied.

The absence of any direct interlayer hopping implies that the number of electrons in each layer
is separately conserved. The corresponding symmetry plays an important role in our analysis.
In contrast, only the {\em total} spin of both layers is conserved.

\begin{figure}
\epsfxsize=2.7in
\centerline{\epsffile{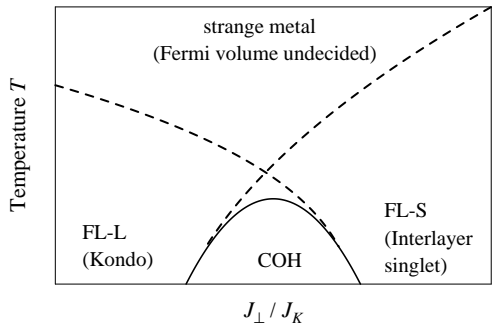}}
\caption{
Schematic phase diagram of the bilayer Kondo lattice model, see text for details.
At $T\!=\!0$ the interlayer coherence phase (COH) breaks a U(1) symmetry; it is
bounded by a finite-temperature Kosterlitz-Thouless transition.
The dashed lines denote crossovers below which the local moments
get screened either by the Kondo effect or by interlayer singlet
formation.
(The in-plane exchange $I$ is assumed to be zero, and magnetic ordering
tendencies due to RKKY interaction, which occur for small $J_K/t$, are ignored.)
\vspace*{-10pt}
}
\label{fig:pd1}
\end{figure}

For situations where the ground state of the system is a Fermi liquid,
it can be characterized by its Fermi volume, i.e., the volume in momentum
space enclosed by the Fermi surface, which is defined by the location of
poles in the conduction electron Green's function at the Fermi level.
The Fermi liquid phase of a single-layer Kondo system is known to
have a Fermi volume, ${\cal V}_{FL}$, counting all electrons in the unit cell, i.e.,
both conduction electrons and local moments,
as determined by the Luttinger theorem \cite{oshi}:
\begin{equation}
{\cal V}_{FL} = {\cal K}_d \, \rho_T\mbox{(mod 2)}. \label{vfl}
\end{equation}
Here ${\cal K}_d = (2 \pi)^d /(2 v_0)$ is a phase space factor,
$v_0$ is the volume of the unit cell of the ground state, $\rho_T
= n_{\ell} + \rho_c$ is the total density of electrons per volume
$v_0$, and $n_{\ell}$ (an integer) is the density of local moments
per volume $v_0$.

For Fermi liquid phases in the present bilayer system, the Fermi volume can be defined separately
for both layers as the number of electrons in each layer is separately conserved.
But what about Luttinger's theorem?
By an extension of the argument in Ref. \onlinecite{oshi}
it is easy to see that only the {\em total} volume of the
Fermi surface of both layers together is constrained. Specifically we have
\begin{equation}
{\cal V}^{(1)}_{FL} + {\cal V}^{(2)}_{FL}  = {\cal K}_d \, \left(\rho^{(1)}_T + \rho^{(2)}_T\right)~~  \mbox{(mod 2)}.
\label{vflbl}
\end{equation}
Here the superscript $1,2$ is a layer index. Thus individual Fermi volumes of the two layers (though well-defined)
are not necessarily required to match the total electron count in their layer.


{\it Phase diagram.}
Let us start with discussing various limits in parameter space.
States without magnetic long-range order can be conveniently discussed keeping $I=0$.
We also assume the conduction band to be away from perfect particle-hole symmetry
to avoid non-generic nesting effects.

For large Kondo coupling, $J_K \gg J_\perp$, each local moment $\vec S_{j\alpha}$
will be Kondo screened by the conduction electrons in layer $\alpha$.
Thus, a heavy Fermi liquid is formed separately in each layer, with a large
Fermi volume per layer -- we dub this FL-L phase.
The effect of a small non-zero $J_\perp$ is to generate spin correlations between the
two Fermi liquids. These will in general lead to innoccuous renormalizations of the
Fermi liquid parameters \cite{note1}.
For dominating interlayer exchange, $J_\perp \gg J_K$, the local moments form
interlayer singlets. For $J_K=0$ the conduction electrons
in each layer form a Fermi liquid with a small Fermi volume per layer (FL-S phase).
Finite $J_K$ can be treated in perturbation theory, leading to spin-spin couplings
of order $J_K^2/J_\perp$ between the two Fermi liquids, which again only lead to unimportant renormalizations
of Fermi liquid parameters \cite{burdin}.

Note that either Fermi liquid satisfies the generalized Luttinger theorem in Eq. (\ref{vflbl}).
Nevertheless the topologies of the Fermi surfaces are quite different.
Indeed it is not possible to smoothly go from FL-S to FL-L without a phase transition.
To see this note that the interlayer exchange symmetry is preserved in both phases:
this symmetry guarantees that the volume (and shape)
of the Fermi surfaces in the two layers are equal.
As the total Fermi volume is quantized according to Eq. (\ref{vflbl}), the individual
Fermi volumes $ {\cal V}^{(1), (2)}_{FL}$ must also be quantized.
However they are quantized to two different values in the two phases. In FL-L,
\begin{equation}
{\cal V}^{(1)}_{FL} = {\cal V}^{(2)}_{FL} = {\cal K}_d (\rho_c + 1)
\end{equation}
while in FL-S
\begin{equation}
{\cal V}^{(1)}_{FL} = {\cal V}^{(2)}_{FL} = {\cal K}_d (\rho_c ) \,.
\end{equation}
Thus the two phases cannot be smoothly connected.

As function of $J_K/J_\perp$ the system evolves from FL-S to FL-L which involves
a change in the quantized value of the {\em Fermi surface volume per layer}: how does this happen?
Besides a direct first-order transition, we show below that intermediate ordered phases which break some symmetry are
very natural. An example is an intermediate phase with spontaneous
interlayer phase coherence (COH), the order parameter being
$\phi = \sum_{j\sigma} \langle c_{j\sigma1}^\dagger c_{j\sigma2}\rangle$.
This phase breaks the separate conservation of electron number
in the individual layers, while preserving total electron number conservation.
An alternate possibility is an interlayer superconductor which
breaks the symmetry of total electron number conservation
while preserving conservation of the {\em difference} of the electron number in the two layers.
In our bilayer system, it does not seem possible to have a direct second-order
``Fermi-volume-changing'' transition between the two Fermi liquids.
Paranthetically, we note that finite exchange interaction $I$ can lead to
various phase with magnetic long-range order -- these will not be considered
further.


\begin{figure}
\epsfxsize=2.7in
\centerline{\epsffile{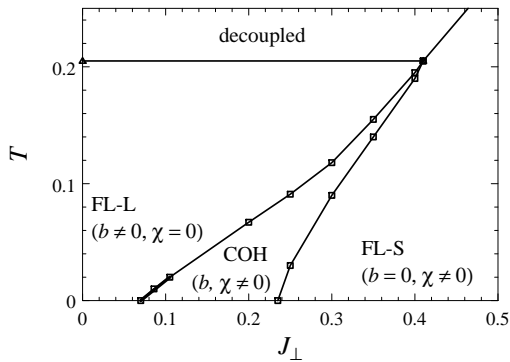}}
\caption{
Mean-field phase diagram, obtained from Eq. (\ref{mf1})
on a bilayer square lattice, with parameters
near-neighbor hopping $t=0.6$, Kondo coupling $J_K = 1$,
conduction band filling $\rho_c = 0.3$, and
a unit cell of one conduction electron site per layer.
Thin (thick) lines denote second (first) order transitions.
Addtional transitions within the COH phase where a Fermi surface sheet
(dis)appears (see text and Fig.~\ref{fig:fsevol}) are not shown.
Only the phase boundaries of the COH phase will survive upon
inclusion of fluctuations around the saddle point.
\vspace*{-10pt}
}
\label{fig:mfpd}
\end{figure}

{\it Mean-field theory.}
We use a fermionic auxiliary-particle representation of the local
moments:
\begin{equation}\label{sf}
\vec S_{j\alpha} = \frac{1}{2} \sum_{\sigma\sigma'}
f^\dagger_{j \sigma\alpha}
\vec{\tau}_{\sigma\sigma'} f_{j\sigma'\alpha}
\end{equation}
where $f_{j\sigma\alpha}$ describes a spinful fermion destruction
operator at site $j$.
As usual, mean-field equations are obtained by decoupling the various
interaction terms, resulting in the mean-field Hamiltonian \cite{hewson}:
\begin{eqnarray}
  H_{\rm mf} & = & \sum_{k\sigma\alpha} \epsilon_k c^{\dagger}_{k\sigma\alpha} c_{k\sigma\alpha} -
  \sum_{j\sigma} \chi_j (f^{\dagger}_{j\sigma 1} f_{j\sigma 2} +
\mbox{
  h.c.} )
    \nonumber\\
  + && \!\!\!\!
       \sum_{j\sigma\alpha} \mu_{f\alpha} f^{\dagger}_{j\sigma\alpha}f_{j\sigma\alpha} -
       \sum_{k\sigma\alpha} b_\alpha
(c^{\dagger}_{k\sigma\alpha}
  f_{k\sigma\alpha} + \mbox{h.c.}) \,.
\label{mf1}
\end{eqnarray}
$\epsilon_k$ is the conduction band dispersion resulting from the $t_{jj'}$,
and we have assumed $I=0$.
The mean-field parameters are determined by the saddle-point equations
\begin{eqnarray}
1 &=& \sum_\sigma \langle f^{\dagger}_{r\sigma\alpha} f_{r\sigma\alpha} \rangle  \,,\\
2 b_\alpha & = & J_K \sum_\sigma \langle c^{\dagger}_{j\sigma\alpha} f_{j\sigma\alpha} \rangle\,, \\
2 \chi_j & = & J_\perp \sum_\sigma \langle f^{\dagger}_{j\sigma 1} f_{j\sigma 2} \rangle \,.
\end{eqnarray}
The above mean-field theory can be justified by generalizing the spin
symmetry from SU(2) to SU($N$) and taking the limit $N\to\infty$ \cite{AffMar}.
(Very similar results are obtained using a Sp($N$) generalization \cite{tJ}.)
Low-energy fluctuations around the mean-field solution are phase
fluctuations of the mean-field order parameters, i.e., local U(1) gauge
fluctuations.

The FL-L phase with Kondo screening is described by non-zero expectation values
$\langle c_{j\alpha}^\dagger f_{j\alpha}\rangle$ at the mean-field level.
The FL-S phase with interlayer singlet formation of the local moments
has non-zero $\langle f_{j1}^\dagger f_{j2}\rangle$.
Note that both quantities involve the auxiliary $f$ particles and
are thus {\em not} physical observables, i.e., they cannot sustain expectation values
once gauge fluctuations are included.

Consider now a state where both $\langle f_{j1}^\dagger f_{j2}\rangle$ and
$\langle c_{j\alpha}^\dagger f_{j\alpha}\rangle$ are non-zero.
Then it follows that $\phi = \langle c_{j\sigma 1}^\dagger c_{j\sigma 2}\rangle \neq 0$ --
this gauge-invariant complex quantity {\em is} a physical observable, which breaks
a physical U(1) symmetry and characterizes a
phase with {\em spontaneous interlayer phase coherence}!
For the two-dimensional system under discussion, small finite temperature will
lead to quasi-long-range order, which disappears in a finite-temperature
Kosterlitz-Thouless transition. No other finite-temperature phase transitions
are expected, as there are no broken symmetries in both FL phases.
The resulting schematic phase diagram is in Fig.~\ref{fig:pd1}.
We have performed a fully self-consistent solution of the mean-field equations
for various sets of parameters,
and a sample mean-field phase diagram is in Fig.~\ref{fig:mfpd}.

\begin{figure}
\epsfxsize=3.5in
\centerline{\epsffile{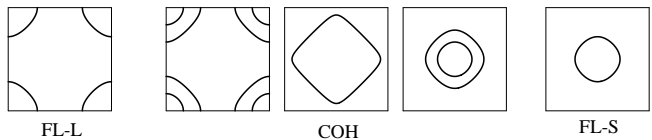}}
\caption{
Schematic evolution of the Fermi surfaces (in two-dimensional momentum space)
along the horizontal axis of Fig.~\ref{fig:pd1},
for a conduction band filling $\rho_c < 1/2$, and $n_{\ell}=1$.
In both the FL-L and FL-S phases there are two identical Fermi surfaces for both
layers, with volumes $1+\rho_c$ and $\rho_c$ respectively.
Upon entering the COH phase, this degeneracy is lifted, and a spontaneous
bilayer splitting develops. Within the COH phase there exist additional transitions
where one Fermi surface sheet (dis)appears; the total Fermi volume (mod 2)
is always given by $2 \rho_c$.
\vspace*{-10pt}
}
\label{fig:fsevol}
\end{figure}

{\it Interlayer coherence phase.}
We now discuss a few properties of the interlayer coherence phase.
It is characterized by the presence of a particle-hole (i.e. excitonic)
condensate, with the expectation value $\phi = \langle c_{j\sigma1}^\dagger c_{j\sigma2}\rangle$
being non-zero, in the absence of explicit interlayer hopping.
Thus there will be a spontaneous bilayer splitting in the electronic
structure, i.e.,
the electrons in both layers will spontaneously form a bonding and an antibonding
band -- this feature should be observable in high-resolution photoemission
experiments.
The COH phase will have an electronic Fermi surface that satisfies the
generalized Luttinger theorem in Eq. (\ref{vflbl}).
However, this Fermi surface will be shared by electrons in the two layers --
in the sense that the quasiparticle states are admixtures of electrons from both
layers.
Thus the notion of two distinct quantized Fermi surfaces associated with either layer
is no longer meaningful.
The Fermi surface evolution from FL-L to FL-S is illustrated in Fig.~\ref{fig:fsevol}.
The broken U(1) symmetry implies a linearly dispersing collective
Goldstone mode with quantum numbers (total) charge 0 and spin 0.

The present interlayer coherence phase is the zero-field analogue of the
corresponding phase in bilayer quantum Hall systems \cite{qhth,qhexp}.
In the quantum Hall context this type of order has also been dubbed
pseudospin ferromagnetism, where the pseudospin degree of freedom
refers to the layer index of a given electron.


{\it Fluctuation effects.}
As argued above, both Fermi liquid phases are stable w.r.t. fluctuations,
their main effect being the restoration of the U(1) gauge symmetry
which is broken at the mean-field level; this also implies the absence of any
finite-temperature transition above the FL phases.

In the interlayer coherence phase amplitude fluctuations of the mean-field
parameters will be gapped. An action for the phase fluctuations can be
formally obtained by integrating out all fermions from the problem;
the result is easily seen to generate the phase mode of the
order parameter with a linear dispersion.


{\it Interlayer superconductivity.}
It is possible to imagine other instabilities near the FL-L -- FL-S transition.
One is the formation of coherence in the particle-particle sector, i.e.,
interlayer superconductivity (with a charge-2 condensate).
At the mean-field level, this can be captured by a mixed decoupling, namely
the $J_K$ term with a particle-hole field $b$
and the $J_\perp$ term with a particle-particle field $\chi$ (or vice versa).
Then, the phase with both $b$ and $\chi$ non-zero will have a non-vanishing
expectation value $\langle c_{j1} c_{j2}\rangle$, representing
interlayer superconducting pairing, i.e., Cooper pairs shared by the two layers,
with $s$-wave orbital symmetry.
We note that the inclusion of Coulomb repulsion in the model will
favor the interlayer coherence phase over the interlayer superconducting phase.


{\it Discussion.}
We have described the physics of bilayer Kondo lattice systems
which contain the competition between the Kondo effect and intermoment exchange.
Two distinct Fermi liquid phases can occur upon varying
the ratio of the Kondo interaction and the interlayer coupling between the local
moments.
These Fermi liquids may be loosely distinguished by whether or not the local moments participate
in the Fermi surface.
A more precise distinction is provided by the topology of their Fermi surfaces
as captured by the notion of the quantized value of the Fermi volume {\em per layer}.
We have argued that a phase with a form of exotic order,
namely spontaneous interlayer coherence in either the particle-hole or the particle-particle channel,
generically intervenes between the two Fermi liquids.
Experimental observation of the former phase may be possible in
angle-resolved photoemission where a spontaneous splitting
between the bands of the two layers
(or a splitting larger than that predicted by band theory)
will be observed.

The finite-temperature properties in the transition region between the two phases
are particularly interesting.
In the region above the phase transition to the ordered phase the behavior is
that of a metal with an undecided Fermi volume per layer, i.e., the
Fermi surface will be strongly fluctuating.
We expect that the properties will be non-Fermi liquid like and possibly similar to that observed
in various ``strange metals'' such as near heavy electron magnetic critical points
or in the optimally doped cuprates.
Interestingly, signatures of fluctuating Fermi surfaces near quantum criticality have been
experimentally detected, e.g., in the heavy electron metal YbRh$_2$Si$_2$ \cite{ybrhsi,silke}
and in the bilayer ruthenate Sr$_3$Ru$_2$O$_7$ \cite{grigera,hooper}.
Developing a theory of strange metal states is a major challenge in condensed matter physics --
we hope that the bilayer systems discussed here can contribute in some way.




We thank S.~Sachdev for illuminating discussions.
This research is supported by the National Science Foundation
grant DMR-0308945 (T.S.) and by the DFG Center for Functional
Nanostructures Karlsruhe (M.V.).  T.S. also acknowledges funding from the
NEC Corporation, the Alfred P. Sloan Foundation, and an award from
The Research Corporation.



\begin{thebibliography}{}
\bibitem{note1} As is usual for any Fermi liquid, at very low temperatures some kind of ordering, such as superconductivity, is
always possible. This will be ignored in our discussion.
\end{thebibliography}
\end{document}